\documentclass[manuscript]{aastex}

\begin{document}

    \title{The Evolution of the Chemical Elements of the Universe }

    \author{Rudolph E. Schild\footnote{Center for Astrophysics,
        60 Garden Street, Cambridge, MA 02138}}
    \date{Received 21 Aug 2007}

\begin{abstract}
Spectroscopic observations of distant cosmological sources continue to
exhibit a surprising result; that the chemical abundance of the universe
seems to be approximately solar for the observed sources at redshifts of 5, 6,
and even 7, even though very few galaxies should have existed at these
epochs and the principal star formation and heavy element production event
should have been at the more local z = 1 - 2.

\end{abstract}


\maketitle

\section{Introduction}

If an informed astronomer would have been asked 20 years ago, what were
the abundances of the elements in galaxies at redshifts z = 5, 6, and 7
(a time when the universe was less than a billion years old, or about 8 \% of
its present age), the reply probably would have been that few galaxies yet
existed and since the inferred age would not permit the first generation of
stars to evolve to the red giant phase in significant numbers, only a few
type II supernovae remnants would have polluted interstellar space with 
heavy elements. Therefore, the metal content would have still been low,
surely less than 10 \% of solar composition and probably less than one
percent. 

Thus the world was surprised when an emerging suite of 8 - 10m class
telescopes with fast spectrographs found that abundances of Lyman-break
galaxies and quasars seem to be greater than solar. For example, the 
1999 oral
report of F. Hamann to the American Astronomical Society (AAS Meeting 
194 - Chicago, Illinois, May/June 1999,
Session 63. ``Evolution of Chemical Abundances over Cosmic Time'') noted that,
"There is a growing consensus from both the emission and intrinsic 
absorption lines that near-QSO environments have roughly solar or higher 
metallicities out to redshifts greater than 4. 
The range is not well known, but 
solar to a few times solar appears to be typical. There is also evidence 
for higher metallicities in more luminous objects, ...."  
At the same special session, Wolfe (1999) noted from studies of 
damped Lyman-alpha
systems that, "Attempts to explain these trends by chemical evolution of
isolated galaxies undergoing passive evolution do not work, since they
predict a systematic increase of the mean $Z/Z_{solar}$ with time, which is not
observed. Rather, the observed trends are better explained in the context
of heirarchical cosmologies by identifying Damped Lyman-alpha systems as
regions of high comoving density in which the mean $Z/Z_{solar}$ rapidly
approaches solar values at high redshifts and saturates thereafter ...." 
Fan et al (2004) report, "The sample of 12 quasars at z 
greater than 5.7 from the SDSS
provides the first opportunity to study the evolution of quasar spectral
properties at $z \sim 6$, less than 1 Gyr after the big bang and only 700 Myr
from the first star formation in the universe. Optical and Infrared
spectroscopy of some $z \sim 6$ quasars already indicate a lack of evolution in
the spectral properties of these luminous quasars; Petrocini et al (2002)
show that the $C_{IV}/N_V$ ratio in two $z \sim 6$ quasars is indicative of
supersolar metallicity in these systems."

The general topic of abundances measured in hi-z galaxies has recently 
been reviewed
by Leitherer (2005). Kewley and Kobulnickey (2005) recently found that 
the O/H ratio is approximately 1/3 of the solar value at z = 3.

The present situation, with measured abundances of high-z sources
higher than expected, is not
thus far viewed as a crisis, because only the most luminous galaxies can
be observed at these redshifts, and their star formation histories may be
atypical. The solar abundances measured in high-redshift quasars are
also not described with alarm, because it is possible to imagine that some
kind of feedback from the central object has influenced the host galaxy's
star formation.

The purpose of the present report is to suggest that the nature of the
baryonic dark matter can probably explain the surprising observational
result. From gravitational lensing of quasars it has been established that
the surprising and rapid microlensing observed indicates that the entire
baryonic dark matter of the universe consists of a population of primordial
planetoids, numbering trillions per galaxy, formed at the time of
recombination. They have since orbited within the halos of all galaxies,
thereby sweeping halos clean of most of the interstellar particles that
would have otherwise contributed to subsequent generations of stars.
Since the role of the baryonic dark matter has thus far not entered the
discussion, it would be surprising if we could fully understand the process of
heavy-element enrichment of the overall universe without considering 
the role of the baryonic dark matter.

In the following paragraphs we examine more carefully the effects of the
primordial planetoid objects and their role in sweeping the interstellar
medium clean of the dust grains deposited by red giants and supernovae.
In section 2 we lay the groundwork for our simple calculation of the physics
of the process. In section 3 we show that the volume swept up by the
primordial planetoids in 100 million years is comparable to the
entire volume of the galaxy, and that it is therefore reasonable to imagine
that the collection of a significant fraction of the interstellar and
circumstellar grains occurred. In section 4 we discuss some implications
and further falsifiable predictions.

\section{The Nature of Primordial Planetoids and their Relationship to the
Interstellar Medium}

Quasar microlensing can reveal the nature of the baryonic dark matter
particles because it has been understood from the outset that the optical
depth to microlensing by stars, or whatever constitutes the baryonic dark
matter, would be so high that observations should show continuous
microlensing. In the case of the discovery object, Q0957+561 A,B, the
optical depth for microlensing in the B image should be 1.35, meaning that
at any statistical moment, 1.35 microlensing events should be underway.
This means that at any statistical moment, probably 1 microlensing event
and a third of the time 2 events, are underway. This would be true if the
physical size of the quasar were nearly the size of the Einstein ring of
the microlensing object, with events having brightness peaks approximately
30 times the un-microlensed brightness. In fact observations show that in
many quasars where microlensing is observed, the amplitudes are much
smaller, which implies that the quasar resolved structure is several times
larger than the solar mass Einstein ring. Schild (1996) found from the
Refsdal-Stabell (1991, 1993, 1997) theory that the Q0957 quasar 
exhibits a luminous surface area 6 times larger
than the area of a statistical half-Solar-mass microlensing star Einstein 
ring.

For star microlenses, the event duration should be about 30 years, with the
event duration decreasing as the square root of the mass of the microlensing
particle. Thus the rapid microlensing detected by Schild (1996) for Q0957
was interpreted to indicate that the mass of the baryonic dark matter
particle was approximately $10^{-5.5} M\odot$. For such microlensing particles,
many million would be along the line of sight to the quasar at any
statistical moment, and a complex pattern of continuous but low-amplitude
microlensing on a one-day time scale would result, as observed 
(Colley and Schild, 2003).

Other explanations have been offered for the microlensing, but none can
reproduce the observations, particularly since microlensing has a unique
observable signature. Because the microlensing takes place in the shear of
strong lensing and star microlensing, a microlens at optical depth near
unity would cause increasing or diminishing brightness with equal
probability, amplitude, and time scale. This has been seen in 
the wavelet analysis of the brightness history of Q0957 (Schild 1999), 
and is a reasonably strong indicator of microlensing. A simulation showing
the rapid microlensing by the primordial planetoids in the presence of
shear of the macrolensing plus star microlensing has been given by Schild
and Vakulik, 2003.

Other manifestations of such a population have since been found in radio
astronomy. The "extreme scattering events" seen in refraction of radio
waves in the direction of quasar Q0954+658 have been associated with
planetary mass clouds which surprisingly seem not to dissipate
(Walker and Wardle, 1998). And strong deflection of radio emission
seen in scintillation observations of pulsar PSR B0834+06 implies the
existence of scattering structures which "may contain a significant
fraction of the mass of the Galaxy" (Hill et al, 2004).

We are of course aware of the claim that MACHO searches have excluded the
existence of a cosmologically significant population of planetary mass dark
matter particles. Of course, the searches do not find significant dark
matter, just the population of white dwarfs always known to exist, and
their non-detections combined with a model of uniformly distributed
(``Gaussian distributed'') planetoids is the basis for the claim 
(Calchi-Novati et al, 2005).  But if
the particles are clumped, as found where they have been detected shadowing
the nebular background with direct HST imaging of the nearest planetary
nebula (Helix), then MACHO searches cannot detect them. Obviously if they are
uniformly distributed, then they are the only gravitating particles in the
universe found unclustered, since stars are found clumped on scales 
of double and
multiple stars, small and large clusters, galaxies, and galaxy clusters.

If the baryonic dark matter is indeed planetary mass particles, as
indicated from quasar microlensing, when did the population form? A
hydrodynamic theory for the formation of primordial structure has been
given by Gibson (1996; see also Gibson \& Schild, 1999) which 
predicted the existence of these
particles at about the time that they were discovered in quasar microlensing.
The particles were formed at the time of recombination, 300,000 years after
the Big Bang, and have collapsed and cooled since. Thus these
objects were orbiting in the halos of galaxies since the first structures
formed, and their sweeping action must presumably have allowed them to
collect up much of the interstellar dust formed in outer atmospheres of
supergiants and supernovae. A supercomputer simulation by Diemand, Moore, and
Stadel (2005) predicts formation in axion dark matter of a primordial 
population of planetary mass particles with the approximate size of the
solar system; if such a process significantly operates in nature, it is
difficult to escape the conclusion that baryonic matter would fall into and
enhance such objects.

The existence of a primordial population of baryonic dark matter primordial
planetoids would have produced two effects that are important 
to the observations
of chemical abundances of galaxies and quasars. First, as noted above, they
would be continually sweeping clean the interstellar medium. To clarify the
description of this, we assume that any typical cubic centimeter volume in
the galaxy collects up some average quantity of metal atoms C at a rate of 
dC/dt, so the total amount swept up over time is Ct. If the test volume is
swept clean every $\tau$ years, then the average metal content is 
$1/2 \times C\tau$. So after the galaxy has been forming stars for 
$\tau$ years, it
would have the time-average mean metal content $1/2 \times C\tau$. 
Since abundances
within galaxies are ordinarily measured and inferred from the emission
lines originating in the galaxy's interstellar medium, sweeping would cause
the appearance of this overall constant average metal abundance. 
Since stars are formed from
these primordial planetoids, successive generations of stars would be
expected to show a more gradually increasing metallicity by this model.

A second important process resulting from the existence of the primordial
planetoid population is the sequestering away of a large fraction of the
hydrogen in the universe, which should have consequences for the
understanding of the cosmological re-ionization event inferred to have 
occurred at z = 6.2. If we assume that at this redshift, $90\%$ of the
hydrogen was bound in these objects, which had condensed from the nearly
uniform primordial gas on a Kelvin-Helmholz time scale of approximately 100
million years, then this gas would not need to be re-ionized for the
universe to become transparent. Therefore the re-ionizing photon flux from
Pop III has been overestimated by a factor of 10.

These two processes combine in an interesting way to give the appearance of
a very rapid increase of the chemical abundances of young galaxies, if the
abundances are estimated from emission lines originating in the
interstellar medium. Notice that before the first sweeping event has
occurred, the interstellar abundance of the typical volume can very quickly
build up to a seemingly large value, because most of the baryonic gas is in
planetoids which do not need to be enriched, and the apparent enrichment of
the galaxy before sweeping in the above example is incorrectly estimated to
be  $10 \times (1/2 \times C\tau)$.

It should not be surprising then that the nature of the baryonic dark
matter would play an important role in defining the observed chemical 
and ionization history and evolution of
the universe. In the following section, we offer a simple calculation to
show that the expected sizes and motions of the primordial planetoids
should be sufficient to significantly sweep up the
particulate matter in the interstellar medium of all galaxies.

As noted above, there must be an important sweeping process tending to
clear the interstellar medium in galaxy discs and halos of the heavy
element ejecta from evolved stars and supernovae. Imagine any cubic parsec
of space in the disc of a galaxy. It is collecting processed gas and dust
for a time short compared to cosmological time. But the polluted cubic
parsec gets swept clean every hundred million years, with the pollutants
heavier than hydrogen and falling to the center of the primordial planetoid
and hence not seen in the kind of spectroscopic measurements a distant
observer makes to study abundances of the interstellar medium. This will
have several consequences for inferences made about cosmo-chemistry by the
terrestrial observer.

To clarify the situation we give a simplified heuristic model of the
process. We assume that all matter in the universe is baryonic, and that a
small fraction $f$ of the primordial gas did not initially condense
into the primordial planetoids and that therefore $1-f$ is that
fraction initially and presently in the form of baryonic dark matter
planetoids (Primordial Fog Particles; Gibson, 1996, 1999). With current
estimates that the universe has approximately 3$\%$ visible matter, we
therefore estimate that $f = .03$ and $1-f = .97$. Note that all 
published calculations
about the emergence of the universe from its dark ages adopt $f = 1.0$, and
$1-f = 0.$

In our scenario, the level of ionization of the universe can be high for
even a weak population of ionizing sources, because the available ionizing
photons are not bothered by the vast reserves of atomic hydrogen
sequestered away in planetoids. A terrestrial astronomer knowing the mean
baryonic (hydrogen) matter density in the universe overestimates the
ionizing flux required to ionize the small fraction $f$ of
interstellar hydrogen. Thus if the primordial planetoids exist,
calculations of the required UV photons can be a factor x30 lower than the
standard calculation, which requires the entire baryonic matter of the
universe to be ionized by z=6. Thus the UV fluxes from quasars are probably
sufficient to ionize the universe, and the existence of a mysterious and
never observed  Population III rich in 300 M$_\odot$ stars is questionable.

This heuristic example also predicts that initially in the universe,
abundances estimated from emission line galaxies are much higher that the
true heavy element abundances, because the primordial planetoids have not
yet swept up much of the heavy elements produced. If, as already adopted in
our example, the true interstellar gas is only 3$\%$ of the baryonic matter,
and will not on average be swept until $10^8$ years after first star
formation, although the interstellar emission lines would indicate solar
metallicity the true baryonic matter enhancement is only 1/30 of solar
metallicity. Thus the process must necessarily cause over-estimation of the
true cosmic enrichment of heavy elements by not correctly taking into
account the nature of the dark matter.

The primordial planetoid population has a third implication for
cosmological evolution of observed abundances. If stars are formed
primarily from the dominant dark planetoid population and not 
from the diffuse
interstellar gas for which we easily measure the emission line strengths
and estimate abundances, then the stars formed will be less enriched than
the observed gas. Again referring to the heuristic example, after $10^8$
years the primordial planetoids have swept galaxy discs and halos once, and
have on average 1/30 of the metals content of the gas; thus stars formed
after $10^8$ years have lower metallicity than the emission line gas.

If we now fast forward in time to the present epoch, we find that the
primordial planetoids play a very different role. If we again accept the
heuristic model and also assume that metals production in the universe has
been nearly constant over cosmic time, and that the interstellar media of
galactic discs and halos are swept clean once in $10^8$ years, then in 3
billion years the metallicity of the diffuse interstellar medium and the
planetoids has equalized. Thereafter, the planetoids have higher
metallicity in the form of cosmic dust collected at their centers.
Subsequent generations of stars that somehow formed from this baryonic 
dark matter now have metallicity greater than that of the galaxy disc
interstellar medium, which is still being swept clean every $10^8$ years.

This picture of star formation, illustrated by a simple heuristic model,
may also illustrate another important aspect of cosmic evolution. The
Gibson (1996) theory does not predict the existence of a primordial
population of stellar mass objects. Instead, it posits that the entire
baryonic dark matter of the universe condensed into a great primordial fog,
with the characteristic mass scale of a fog droplet estimated at $10^{-7} 
M_{\odot}$, and observationally estimated from quasar microlensing 
(Schild, 1996) to be $10^{-6} M_{\odot}$. Thus the first stars were made 
from accretional cascade when the
stickey planetary mass ``Primordial Fog Particles'' interacted at their
fuzzy boundaries to form pairs, and then pairs of pairs, etc. It is likely
that such an accretional cascade would have produced objects with
a mass function that is strongly biased toward low masses. Thus the early
universe should have been dominated by an initial population of low-mass
and low-luminosity objects with Pop III chemical composition. Of course it is
possible that any available gas also formed Pop III stars from interstellar
gas by the more accepted modern theory of star formation. By this argument
it is likely that the Pop III initial mass function is different from the
present epoch stellar mass function.

\section{An Estimate of the Sweeping Activity of Primordial Planetoids}

In the following paragraphs, we estimate the total sweeping action by the
primordial planetoids from reasonable estimates of the planetoid cross
section area and the speed. We assume a population of primordial objects
orbiting our Galaxy with a mean motion of 220 km/sec. In principle the
cross section area of such an object should be derived from a complex model
of a brown dwarf-like object of terrestrial mass and primordial gas
composition, formed at the time of recombination in a condensation-void
separation process as envisaged by Gibson (1996), and cooled through the 
13-billion year history of the universe. The complex processes of cooling
and external re-heating have already been discussed by Schild and Dekker
(2006) with the conclusion that it is better to establish a reasonable
diameter estimate from observations. Objects of the expected type have been
identified and discussed by Gibson and Schild (2003) residing in the Helix
planetary nebula and seen as "Cometary knots". Although these objects have
been lightly dismissed as Rayleigh-Taylor gas instabilities, their high
density contrast with the ambient gas makes such an explanation
implausible (Gibson and Schild, 2003; see sections 2.3.1 and 4)
and no detailed supporting instability model exists. Nor can 
they be understood as shock instabilities, since they do not exhibit a 
shock spectrum. 

Thus we take these cometary knots to be remnant primordial planetoids,
labeled Primordial Fog Particles by their discoverer (Gibson 1996) and we
take their sizes to be as measured from direct imaging by Meaburn
et al (1998). Since their outer
atmospheres have been ablated back and away, only the lower limit to their
sizes can be estimated from the observations; this limit is taken to be
$10^{15} cm$. 

For this diameter, the cross-sectional area for sweeping will be $10^{30}
cm^{2}$ and the distance travelled in $10^{9}$ years would be 
$10^{24} cm$, for a
total swept volume of $10^{54} cm^3$ per planetoid. For a planetoid mass of
$10^{-6} M\odot$ and total Galaxy mass of $10^{11} M\odot$, there would be 
$10^{17}$ of the
sweeping particles for a total swept volume of $10^{71} cm^3$. For a Galaxy
Halo diameter of 50 kpc, the volume is $10^{70} cm^3$, implying any statistical
small volume of space in the Galaxy or Halo gets swept 10 times in a
billion years. Thus the constant $\tau$ introduced in the previous section is
estimated to be 100 million years.

In this calculation we have assumed that any dust grain or molecule
encountered by the primordial planetoid is captured. It is of course
unknown to what extent this is true. Lacking real information about the
structure of the outer atmosphere of such a planetoid, we examine the
properties of meteors seen in the terrestrial atmosphere.

It is reasonably well known that meteors swept up by the Earth's atmosphere
are observed glowing at altitudes near 100 km (Millman and McKinley, 1963). 
The brightest objects, presumably the more massive entering
particles, are observed penetrating to 70 km, where the atmospheric density
is $10^{-7} g/cm^3$. However their firey trails begin at an altitude where
the smallest meteors occur, near 100 km, with a residual atmospheric
density of $10^{-9} g/cm^3$. Presumably the smaller, finer 1 micron sized
particles can be captured at higher altitudes with lower densities; recall
that spacecraft operate in low Earth orbit near 160 km, where the residual
density is $10^{-12} g/cm^3$. But although low orbiting spacecraft survive
for multiple orbits because of their large mass/surface area ratio, the
interstellar dust particles would not. 

\section{Discussion About Further Properties of the Baryonic Dark Matter}

The proposed role of the primordial planetoids in sweeping the interstellar
medium throughout the history of the universe must have further
implications for the observed chemical evolution of the universe.
It is easy to understand that insofar as the primordial planetoids maintain
a relatively constant sweeping action throughout the discs and halos of
galaxies, the rate of accumulation of heavy elements produced in red giants
and supernovae in the gaseous interstellar medium is diminished and many of
the heavy elements must be sequestered away until the planetoids contribute
to star formation at a later time. It must be true to some extent
that star formation causes the conversion of dark baryonic matter into 
visible matter, but that stars become fading white dwarfs after their hydrogen
burning lifetimes, to again become dark matter. The process must convert
pristine primordial halo dark matter into heavy element enriched 
dark degenerate cores in discs of spiral galaxies. It must also sequester
away heavy elements out of the interstellar medium but yield the heavy
elements up again in later generations of stars being formed out of the
primordial planetoids.

Thus as the primordial planetoids are acting as an important reservoir,
sequestering away heavy elements for a future generation of stars, they
must have a complex life of their own. The hydrodynamic theory that
predicts their formation (Gibson 1996,) also comments about more
specific processes occurring in their formation and evolution. Following
their formation in a condensation/void separation process, with masses
top-limited at terrestrial mass, $10^{-6} M\odot$, the early objects would have
slowly collapsed on a Kelvin-Helmholtz time scale of 300 million years but
must have been quite sticky for this duration, since their extended
residual atmospheres would probably have created large viscosities in the
extended outer parts. Thus 2-body interactions should have caused mergers,
and these merger products would presumably have caused small departures from
the ambient flow, inducing further mergers in an unstable cascade process.
Although the available computer simulations of the process show the
formation of the condensations (Truelove 1998) \footnote{ The Truelove
report is written in the context of Jeans theory, and starts with the
presumption that instabilities to condensation found in the simulations
must be some artifact of the calculations. We think it more likely that the
simulations are correct, and that quiescent gas is unstable to the
condensation-void separation process, if other fluid forces are not
disruptive.} 
the calculations become
unstable after the formation of first condensations and have not been
continued with finer sampling to follow the expected interaction process.
We presume that the process is intrinsically unstable for the smooth
expanding flow of the primordial gas, and that a cascade of merging
Primordial Fog Particles, (PFP's Gibson 1996) would generate a population
of merged primordial objects and thereby extend the mass range much higher
than the viscosity-limited primordial cutoff. Thus the primordial process
of condensation-void separation should produce cascades to heavier masses
of the primordial planetoids, which will also exhibit the observed
clumping in space.

The Gibson (1996) theory describing this "Primordial Fog Particle"
formation also gives nod to the so-called Jeans mass scale, which then
predicts that at the time of recombination, as the primordial planetoids
were forming, mass was also aggregating on globular cluster mass scales,
$10^{6} M\odot$. Since these form at the same gas 
density as the planetoids, they
would have comparable Kelvin-Helmholz time scales for collapse. Since the
primordial planetoids within such proto-globular-clusters thus become the
particles of a gas ultimately establishing a virialized structure, the
relative motions of the primordial planetoids and hence the merging to
higher mass would be greater within the proto-globular-clusters.
Thus the merging of the
primordial planetoids and the process of first star formation from them,
both proceeding on a time scale of 300 million years, would be different
within the cluster, and would presumably result in a difference in the
population III mass spectrum inside/outside the cluster. It also seems
likely that the first Population III star formation would have occurred in
the Gibson (1996) cluster-mass condensations. 

The formation of the primordial planetoids must also importantly affect
the amount of ionizing radiation needed to re-ionize the universe following
the "dark ages". For the universe to become transparent to radiation
shortward of the Lyman limit, only the fraction of gas that remains
following the efficient condensation-void separation, needs to be ionized.
Given that the universe is presumed to have ten times more baryonic
dark matter than
luminous matter today, and that the dark fraction must have been larger
before the star formation peak at z = 1 - 2, we can estimate that only about
a percent of the baryonic matter in the universe needed to be ionized for
the universe to become transparent, reducing the required amount of ionizing
radiation from the first generation of stars and from the quasars by a
factor of at least one hundred. Current estimates for the required
ionization ultraviolet flux by Wyithe and Loeb (2003) do not take into
account the reduction in required radiation implied by the condensation of
this dominant dark matter population.

The more recent history of the primordial planetoids is of further interest
because in addition to their carrying a large fraction of the heavy metals,
they must also figure importantly in the formation of the solid body
objects of the solar system (planets, moons, comets, asteroids, and Kuiper
Belt Objects). Recall that the presently accepted formation picture 
involves rocks
colliding with rocks and sticking together over a vast range of phase
space, even though this process has never been observed to occur in the
history of our civilization. And when the New York World Trade Center
fell in 2001, nobody reported that all the rocks were stuck together
in the bottom of the hole.

An alternative view is that as the primordial planetoids swept the
interstellar medium clean of dust, most of it 
probably fell to the center, or was
vaporized, only to crystallize out on other nucleation centers at a later
time. But the planetoids would have undergone several phase changes, first
to liquid and then to solid state as they cooled to the present 2.7 K
temperature of the universe. And even after freezing, further 
contraction of the cooling object would imply crushing central forces. 
This has almost certainly created the dusty/gassy cores of comets, and the 
surprising result that comets seem to be made of compacted dust, not rocks
(a surprising conclusion from the NASA impact of a 372 kg projectile with
comet Temple 1 (Tytell, 2005)). \footnote{ Early reports quoted from
project scientists by author Tytell stress the surprising conclusion that, 
"The microscopic particles of dust are not microscopic because
of the impact; it's just that we released microscopic particles," says
[project scientist] McFadden. "In other words, the impact didn't pulverize
the dust; the particles were small to begin with.''}

Recall also that the theory
justifying the use of 90 HST orbits to find an expected 60 faint
Kuiper Belt objects found only
3, showing that the collision theory for the creation of Kuiper Belt
Objects was flawed. Thus G. Bernstein reported in HST Newsletter v21 No.1 p18
Winter 2004 that, "Our search for TNOs was spectacularly unproductive; we
discovered only 3 TNOs - nearly 30 times fewer than expected by
extrapolations from brighter surveys."
We predict that the KBOs will have compacted dust
composition like the comets, and that they exist in the outer fringes of
the solar system as the compacted core solid remnants of a fraction of
the million primordial planetoids whose gas was deposited in the forming 
sun but whose compacted dust solid cores were left behind in distant orbit.


\begin{thebibliography}{}

\bibitem[cn05]{cn05} Calchi-Novati, S. et al, 2005, \aap, 443, 911

\bibitem[cs03]{CS03} Colley, W. \& Schild, R. 2003, \apj, 594, 97

\bibitem[ddt]{d05} Diemand, J. Moore, B. \& Stadel, J. 2005, NATURE, 
  433, 389 

\bibitem[f04]{F04} Fan, X. et al, 2004, \apj, 128, 515

\bibitem[2003]{Gib03} Gibson, C. 1996, Appl. Mech. Rev., 49, 299;
astro-ph/9904260

\bibitem[gs99]{gs99} Gibson, C. \& Schild, R. 1999, astro-ph/9904362

\bibitem[gs03]{gs03} Gibson, C. \& Schild, R. 2003, astro-ph/0306467

\bibitem[h99]{h99} Hamann, F. 1999, BAAS, 31, 924

\bibitem[hs04]{HS04} Hill, A. et al, 2005, \apj, 619, L171

\bibitem[kk05]{kk05} Kewley, L. \& Kobulnickey, H. 2005, in Starbursts:
From 30 Doradus to Lyman Break Galaxies, ed. R. DeGrijs and R. Gonzales
Delgado [Dordrecht:Springer] p. 307

\bibitem[l05]{l05} Leitherer, C. 2005, "Metals in Star-forming Galaxies at
High Redshift," review paper in Proceedings IAU Symposium 228, ed. V. Hill,
P. Francois, \& F. Primas, p. 551

\bibitem[Meab]{M98} Meaburn, J. et al, 1998, Monthly Not. R.A.S. 294, 201

\bibitem[mm]{mm} Millman, P. \& McKinley, D. 1963, in The Moon, 
Meteorites, and Comets, ed B. Middlehurst and G. Kuiper [Chicago:
University of Chicago Press] p. 674 

\bibitem[1991]{RS91} Refsdal, S., and Stabell, R. 1991, \aap, 250, 62

\bibitem[1993]{RS93} Refsdal, S., and Stabell, R. 1993, \aap, 278, L5

\bibitem[1997]{RS97} Refsdal, S., and Stabell, R. 1997, \aap, 325, 877

\bibitem[1996]{rs96}Schild, R. 1996, ApJ, 464, 125

\bibitem[rs99]{rs99}Schild, R. 1999, ApJ, 514, 598

\bibitem[1996]{sd05}Schild, R. \& Dekker, M. 2006, A.N. 327, 729

\bibitem[2003]{SV03} Schild, R. E., \& Vakulik, V., 2003, \aj, 126, 689

\bibitem[1998]{t98} Truelove, J.K. et al, 1998, \apj, 489, L179

\bibitem[t05]{ST05} Tytell, D. 2005 Sky and Telescope, 110, 34

\bibitem[ww98]{ww98} Walker, M. \& Wardle, M. 1998, ApJ, 498, L25

\bibitem[w99]{w99} Wolfe, A. 1999, BAAS, 31, 924

\bibitem[ws03]{WL03} Wyithe, S. \& Loeb, A. 2003, IAU Symposium 216, p.154

\end{thebibliography}
\end{document}